\documentclass[twocolumn,appendixfloats,tighten]{aastex6}
\usepackage{newtxtext,newtxmath}
\usepackage[T1]{fontenc}
\usepackage{ae,aecompl}
\usepackage{amsmath}	
\usepackage{amssymb}	
\usepackage{mathtools}
\usepackage{ctable}
\usepackage{url}
\usepackage{xspace}
\usepackage[normalem]{ulem}
\usepackage{hyperref}
\usepackage[all]{hypcap}
\usepackage{graphicx}
\usepackage{color}
\usepackage{float}

\def\msun{{\rm\,M_\odot}}

\newcommand{\dd}{\ensuremath{\mathrm{d}}}
\newcommand{\diff}[2]{\ensuremath{\frac{\dd #1}{\dd #2}}}

\def\msun{{\rm\,M_\odot}}

\newcommand{\be}{\begin{equation}}
\newcommand{\ee}{\end{equation}}

\newcommand{\chieff}{\ensuremath{\chi_{\rm eff}}\,}
\newcommand{\chieffbar}{\ensuremath{\bar{\chi}_{\mathrm{eff}}}}
\newcommand{\sigmachieff}{\sigma_{\chi_\mathrm{eff}}}
\def\h2{${\rm\,H_2}$}

\usepackage{color}

\begin{document}

\title{A trend in the effective spin distribution of LIGO binary black holes with mass}

\author{Mohammadtaher Safarzadeh\altaffilmark{1}, Will M. Farr\altaffilmark{2,3}, and Enrico Ramirez-Ruiz\altaffilmark{1,4}}

\altaffiltext{1}{Department of Astronomy and Astrophysics, University of California, Santa Cruz, CA 95064, USA \href{mailto:msafarza@ucsc.edu}{msafarza@ucsc.edu}}
\altaffiltext{2}{Department of Physics and Astronomy, Stony Brook University, Stony Brook, NY 11794, USA}
\altaffiltext{3}{Center for Computational Astronomy, Flatiron Institute, New York, NY 10010, USA}
\altaffiltext{4}{Niels Bohr Institute, Blegdamsvej 17, 2100 K\o benhavn \O, Denmark}

\begin{abstract} Binary black holes (BBHs) detected by gravitational wave (GW)
observations could be broadly divided into two formation channels: those formed
through field binary evolution and those assembled dynamically in dense stellar
systems. Each of these formation channels, and their sub-channels, populate a
distinct region in the effective spin-mass (\chieff$-M$) plane.   Depending on
the branching ratio of different channels, an ensemble of BBHs could show a
trend in this plane. Here we fit a mass-dependent distribution for $\chieff$ to
the GWTC-1 BBHs from the first and second observing runs of Advanced LIGO and
Advanced Virgo.  We find a negative correlation between mass and the mean
effective spin ($\chieffbar$), and positive correlation with its dispersion
($\sigmachieff$) at 75\% and 80\% confidence. This trend is robust against the
choice of mass variable, but most pronounced when the mass variable is taken to
be the chirp mass of the binary. The result is consistent with  significant
contributions from both dynamically assembled and field binaries in the GWTC-1
catalog. The upcoming LIGO O3a data release will critically test this
interpretation. \end{abstract}

\section{Introduction}

The spin probability distribution function of stellar mass black holes at birth
is unknown. The efficiency of angular momentum (AM) transfer from the core of a
dying star to outer shell layers through magnetic fields sets the expected spin
of the newly born compact object. The efficiency of the mechanism is debated in
the literature, for example, the Geneva stellar evolution model
\citep{Eggenberger:2007dl,Ekstrom:2011ke} assumes moderate efficiency of AM
transport through meridional currents and therefore permits BHs to be born with
non-negligable spin, while efficient transport by the Tayler-Spruit magnetic
dynamo \citep{Spruit:1999vt,Spruit:2001ki}, as implemented in stellar evolution
calculations \citep{Fuller:2019gc,Fuller:2019jz} predicts all isolated BHs to be born
very slowly rotating.

The models in which BHs are assumed to be non-spinning at birth have a difficult
time explaining the observed high spin of the BHs in high mass x-ray binaries
(HMXBs) \citep{2017ApJ...846L..15B, Qin:2018cl} since these black holes are wind
fed and therefore gas accretion \citep{Fragos:2014wz} or tidal locking
\citep{Zaldarriaga:2017fn,2018ApJ...862L...3S}  can not explain their high
spins. Moreover, one of the BHs in GW151226 has spin greater than 0.2
\citep{Abbottetal:2016ez} which might challenge the zero spin scenario. Given
that the BHs in high mass x-ray binaries are less massive compared to the LIGO
BHs, it is possible there may be a mass trend for the spin of the black holes.
In addition to the effects of angular momentum transport discussed above, such a
trend could be due to the supernova explosion mechanism that form the BHs or
secondary astrophysical mechanisms such as tidal locking
\citep{Zaldarriaga:2017fn,2018ApJ...862L...3S} or gas accretion that can change
the spin of a BH \citep{Fragos:2014wz}.  This latter mechanism is operative in
the case of low mass x-ray binaries (LMXBs).

The effective spin of a binary black hole (BBH) system is defined as
\be
\chi_{\rm eff}\equiv\frac{m_1 a_1 \cos(\theta_1) +m_2  a_2  \cos(\theta_2)}{m_1+m_2} ,
\ee
 where $m_1$, and $m_2$ are the masses of the primary and secondary black hole, and $a_1$, and $a_2$ their associated dimensionless spin magnitude defined as:
 \be
 a=\frac{c J_{\rm BH}}{G M_{\rm BH}^2}.
 \ee
 Here $c$ is the speed of light, G is the gravitational constant, and M and J are  the mass and angular momentum of the BH.
$\theta$ is the angle between the direction of each BH's spin and the orbital angular momentum of the BBH.  The effective spin parameter is the best-measured spin-related parameter from gravitational wave observations \citep[][and references therein]{2017Natur.548..426F}, so here we focus on this one-dimensional summary of the full, six-dimensional space of BBH spins.

The spin distribution of the LIGO black holes therefore carries crucial
information that illuminates the formation process of these systems \citep{2017CQGra..34cLT01V,2017Natur.548..426F,2017MNRAS.471.2801S}. Broadly,
two different mechanism have been proposed for the formation of the BBHs: (i)
assembled in the field through stellar evolution and a potential common
envelope phase, (ii) assembled dynamically, either in globular or nuclear star clusters or hierarchical triple or higher order stellar systems
\citep{Rodriguez:2018bs,2019arXiv191104495S}. Each of these channels predict a
different spin-mass distribution: Field binaries are expected have their BH
spins preferentially aligned with the orbital angular momentum of the binary
\citep{Belczynski:2002gi,Dominik:2012cwa,Zaldarriaga:2017fn,Gerosa:2018hw,Qin:2018cl,Bavera:2019ut,2018ApJ...862L...3S},
 while dynamically assembled binaries
\citep{Zwart:2004jj,2014ApJ...784...71S,Chatterjee:2016fl,Rodriguez:2016hi,Antonini:2017el,2018ApJ...853..140S,Rodriguez:2018ci}
are expected to have their spin isotropically distributed with respect to the angular
momentum of the binary and therefore result in large fraction of the systems to have \chieff$ <0$.

The effective spin parameter for the 10 LIGO/Virgo GWTC-1 BBHs is consistent with being clustered all around zero \citep{Abbottetal:2018vb,Belczynski:2017wa,Roulet:2019js} which could be due the fact that
LIGO black holes are mostly non-spinning or their spins lie in the orbital plane of the binary.  Here we examine this population for mass-dependent effects on effective spin.

Several additional BBH merger events have been claimed in the same LIGO/Virgo
data used to generate GWTC-1 \citep{Venumadhav2019,Venumadhav2019b}, including
some with (large) positive \citep{Zackay2019} and negative effective spin
\citep{Venumadhav2019b}.  \citet{Piran2019} argue that the larger catalog is
more consistent with field than dynamical formation (using models where the
entire population comes from a single channel).  Here we focus only on the BBH
systems in GWTC-1 for two reasons: (1) full posterior distributions for the
parameters of the additional events have not been made available, and a Gaussian
approximation to the $\chieff$ posterior may not be adequate for population
analysis in such a large catalog \citep{Ng2018} and (2) there is no
publicly-released procedure to characterize the sensitivity of the pipelines
used in \citet{Venumadhav2019,Venumadhav2019b} well enough to account for
selection biases in the population.

We summarize how different formation channels of the BBHs populate different
regions in \chieff-mass plane in \S \ref{sec:sec2}. In \S \ref{sec:method} we
analyze the joint mass-effective spin distribution of the ten LIGO/Virgo BBHs to
search for possible correlations of the mean and dispersion of the effective
spin with mass, where mass can be either the primary mass, the chirp mass, or
the total mass of the binary. In \S \ref{sec:summary} we summarize our results
and suggest alternative joint distribution studies that could carry similar
information as \chieff-mass distribution.

\section{Summary of distribution of BBHs in effective spin-mass plane}\label{sec:sec2}

\begin{figure*}
\includegraphics[width=1.0\linewidth]{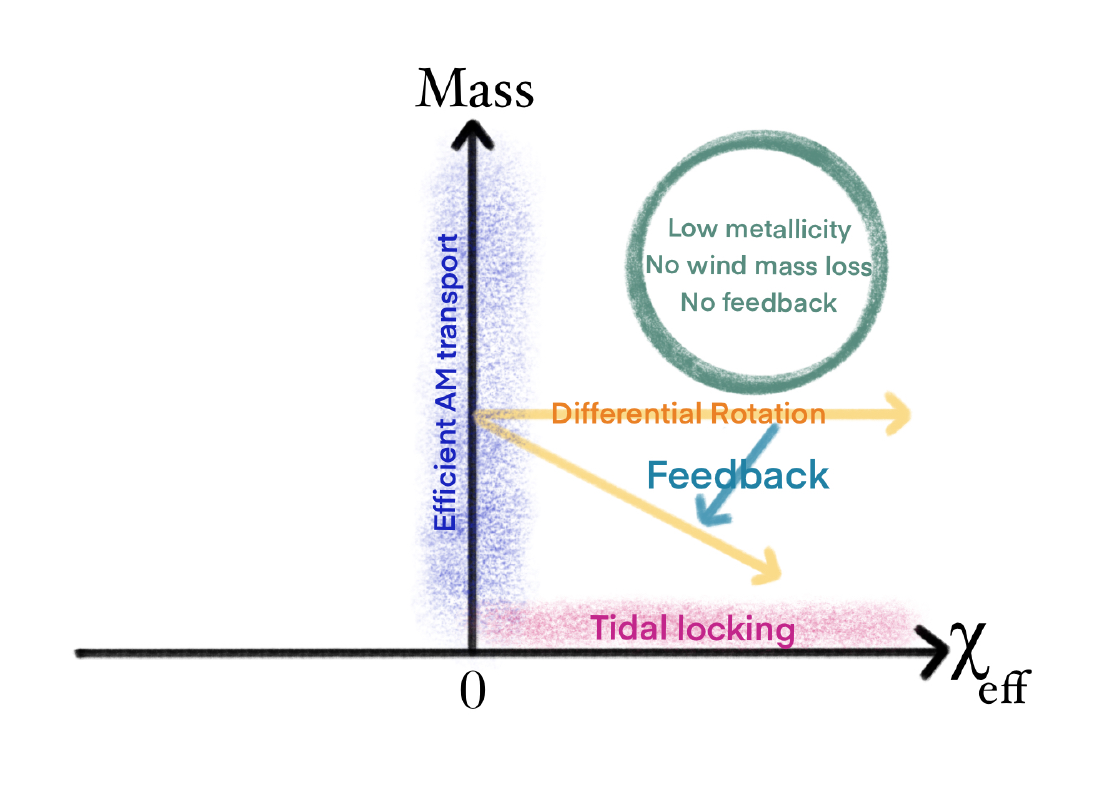}
\caption{The expected distribution of black holes in spin-mass space based on different assumptions regarding the formation of the black holes at birth, and their subsequent evolution in the presence of
secondary astrophysical mechanisms. The blue shaded band shows models in which high efficiency of angular momentum transport leads to formation of almost non-spinning black holes \citep{Fuller:2019gc,Fuller:2019jz}. The leakage into negative \chieff could arise from natal kicks of the BHs upon formation that result in spin-orbit misalignment \citep[][and references therein]{OShaughnessy:2017gk}.
The green circle shows the result of moderate angular momentum transport at low metallicities where wind mass loss is quenched \citep{Kudritzki:2000cp,Vink:2001cs}
and therefore it is possible to form massive spinning systems.
The pink region indicates a secondary mechanism that leads to formation of high spin systems at low masses through tidal interactions,
where the secondary star is spun up due to synchronization with the orbiting companion BH before collapse \citep{Zaldarriaga:2017fna,Gerosa:2018hw,Bavera:2019ut}.
The arrow depicts the expected distribution when the feedback from compact object formation is taken into account. In this model, depending on the initial rotation of the star, a disk can form whose feedback prevents
the collapse of outer stellar layers on to the compact object which results in lower mass BHs with lower spin parameters \citep{Batta:2019vd}. Except the pink region, the other three regions show the primary source of spin in the newly born BHs before being spun up in tidal interactions.}
\label{f:origin}
\end{figure*}

Different models for the evolution of field binaries predict different \chieff-mass distribution:
There are models that predict all isolated black holes should be born slowly rotating \citep[e.g.,][]{Fuller:2019gc,Fuller:2019jz} and therefore, secondary astrophysical mechanisms
such as tidal interactions are invoked to explain fast rotating BHs
in either HMXBs or GW151226. Effectively such models predict a distribution of BHs in mass-spin at birth similar to the blue band in Figure \ref{f:origin} for the BBHs.

In the case of moderate efficiency of AM as implemented in MESA stellar evolution model \citep{Eggenberger:2007dl}, low metallicity stars are expected to not lose mass through winds, as their opacity for EUV/UV photons is small \citep{Kudritzki:2000cp,Vink:2001cs}.
Therefore, the collapse of such stars is expected to result in both massive, and highly spinning BHs \citep[although in such cases feedback is likely to limit their mass;][]{Batta:2019vd}. The predicted locus of such objects are depicted by the green circle in Figure \ref{f:origin}.
Since low metallicity environments are
thought to be the underlying requirement for the formation of such BBHs, given the metallicity evolution of the universe, these systems are expected to be likely born at high redshifts (although it is possible to
form such systems in pockets of the low metallicity regions in the local universe), and therefore, a long delay time (large separations at birth) are thought to make them merge at $z<0.2$ such that LIGO can see them.
This channel has been proposed to explain GW170729 \citep{Bavera:2019ut}.

Other models with inefficient angular momentum transport can produce a
distribution in $\chieff$ space with a mean value that decreases and a
dispersion that increases with increasing mass \citep{Belczynski:2017wa}, as we
find in the GWTC-1 catalog.  However, such models generally predict higher
values of $\chieff$ at low masses than are observed in GWTC-1, and are therefore
disfavored \citep{Belczynski:2017wa}.  Nevertheless, it may be possible to
produce the trends we observe in mean and dispersion with mass through field
formation models invoking a combination of efficient angular momentum transport,
tidal spin-up, and metallicity-dependent mass loss from high-mass stellar winds
that differs from the one explored in \citet{Bavera:2019ut}.

Dynamical assembly can also make massive spinning BBH mergers, however, the expected distribution in the case of dynamical assembly
is symmetric and therefore future BBH detections can tell us whether there is a locus of BBHs at high mass and high spin, or whether the massive BBHs are symmetrically distributed in \chieff.
This can shed light on the underlying formation mechanism of such systems and whether their spins have been altered by subsequent stellar encounters \citep[e.g.,][]{2019ApJ...877...56L}.

Under dynamical assembly, the predicted \chieff-mass distribution is more well
defined in its structure: first generation black holes could be born with zero
spin, however the final merger product of such black holes will have high spins
($a\approx0.7$) \citep{2017ApJ...840L..24F,2017PhRvD..95l4046G}. The merger of
these second generation BHs with either another second generation BHs, or first
generation BHs, forms a BBH with at least one of the BHs to be highly spinning.
The spin orientation of the BHs in the dynamical assembly would be random with
respect to the binary's orbital plane and therefore the final merger product is
expected to show a symmetric \chieff distribution around zero that widens at
larger masses. The widening at larger masses is due to the merger products of
higher generation black holes in a dense cluster like environment
\citep{Rodriguez:2019fx,Doctor:2019ub}.

\begin{figure*}
\includegraphics[width=1.0\linewidth]{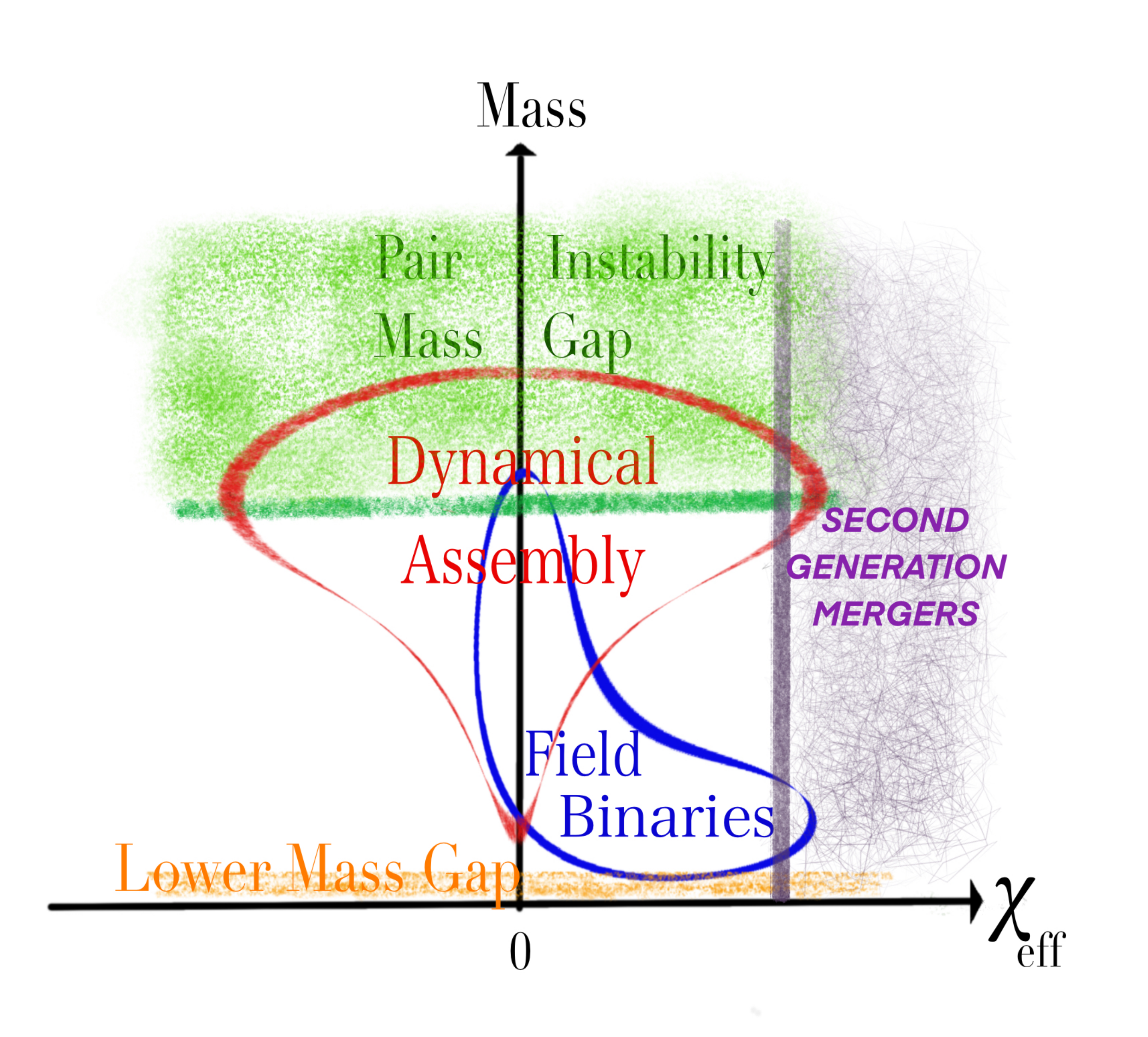}
\caption{The expected distribution of black holes in spin-mass space from dynamical assembly (red pear like region), and from field formation (blue banana like region).
These two regions demonstrate the broad sketch of how two main formation channels would occupy the \chieff-mass space. The debated lower mass gap, and pair instability mass gap are shown with orange and green
bands respectively (see text for brief explanation of the two bands). The purple region indicates systems with $\chieff>0.7$ where second generation mergers become important for the dynamical scenario. The dynamical assembly can assemble BBHs that exceed the PISN mass gap. The field binaries can provide the negative trend with mass, and the dynamically assembled binaries provide the increase in dispersion with mass, the combination of which can potentially explain our findings.}
\label{f:origin_combined}
\end{figure*}

Figure \ref{f:origin_combined} shows the expected distribution in \chieff-mass from two separate categories: Field binaries shown as the blue banana region, and the dynamically assembled
binaries shown with red pear like region. These are rough sketches of the expected distributions and not necessarily to scale. The lower orange band indicates the debated lower mass gap (between 2-5 $\msun$).
The green band and the light green region above it indicates the presence of a drop in mass function of the BBHs given the 10 LIGO/Virgo BBHs \citep{Fishbach:2017ic,Talbot:2018cj,Roulet:2019js,Safarzadeh:2019fh} due to
pulsational pair instability supernovae \citep{Woosley:2017dj} where BHs with mass between $\approx50-150 \msun$ are expected to not form.
The field binaries can provide the negative trend with mass, and the dynamically assembled binaries provide the increase in dispersion with mass, the combination of which can potentially explain our findings.

\section{Regression analysis on GWTC-1}\label{sec:method}

In this section we analyze the joint \chieff-mass distribution of the ten LIGO/Virgo BBHs to search for possible correlations of the mean and dispersion of the
effective spin with mass, where mass can be either the primary mass, the chirp mass, or the total mass of the binary.

We assume that the population distribution for $\chi_\mathrm{eff}$, conditioned
on mass, follows
\begin{equation}
  p\left(\chi_\mathrm{eff} \mid m \right) = {\mathcal N}\left( \mu\left(m\right), \sigma\left( m \right) \right) T[-1, 1],
\end{equation}
where ${\mathcal N}\left( \mu, \sigma \right)$ is a normal distribution with mean $\mu$, and dispersion $\sigma$. The notation $T[-1,1]$ means to truncate the normal distribution to the range $[-1, 1]$.
The parameters $\mu$ and $\sigma$ are set by the mass of the BBH via
\begin{equation}
  \mu(m) = \mu_0 + \alpha \times \left( \frac{m}{30 \, M_\odot} - 1 \right),
\end{equation}
and
\begin{equation}
  \sigma(m) = \sigma_0 \exp\left(\beta\times \left( \frac{m}{30 \, M_\odot} - 1 \right) \right).
\end{equation}
Here $m$ can be any measure of the mass of the BBHs: the primary mass of the BBH
($m_1$), the total mass of the system ($M_{\rm tot}=m_1+m_2$), or the chirp mass
of the system,
\be
M_{\rm chirp}= \eta^{3/5} M_{\rm tot},
\ee
where $\eta$ is the symmetric mass ratio given by
\be
\eta=\frac{m_1 m_2}{(m_1+m_2)^2}.
\ee
We assume a fixed mass and redshift distribution consistent with the population
analysis in \citet{Abbottetal:2018vb} throughout this work, with
\begin{eqnarray}
  p\left( m_1 \right) & \propto &  \frac{1}{m_1}, \label{eq:m1-pop} \\
  p\left( m_2 \mid m_1 \right) & \propto & \mathrm{const} \label{eq:m2-pop} \\
  p\left( z \right) & \propto & \left( 1 + z \right)^{1.7} \diff{V}{z}, \label{eq:z-pop}
\end{eqnarray}
where $V(z)$ is the comoving volume \citep{1999astro.ph..5116H}.  (The redshift
distribution gives an observed merger rate that is consistent with the comoving
merger rate tracking the low-redshift evolution of the star formation rate
\citep{Fishbach:2018,Madau:2014gt}.)  We have verified that our results do not
depend on the choice of mass and redshift distribution within the ranges
permitted by the population analysis of \citet{Abbottetal:2018vb}.

The marginal likelihood for the catalog data $d_{GW}$ given population parameters $\Theta \equiv\left(\mu_0,\sigma_0,\alpha,\beta\right)$ is \citep{Mandel:2018jd}
\begin{multline}
  \label{eq:catalog-likelihood}
p\left( d_{GW}\mid \Theta \right) =
 \\ \prod_{i=1}^{N} \frac{1}{\alpha\left( \Theta \right)} \int \dd \chi_\mathrm{eff}^{(i)} \, \dd m^{(i)}_1 \, \dd  m^{(i)}_2 \, \dd z \, \\ \times  p\left( d_{GW}^{(i)} \mid \chi_\mathrm{eff}^{(i)}, m_1, m_2, z \right)
 \\ \times p\left(\chi_\mathrm{eff}^{(i)} \mid \Theta,m_1, m_2 \right) p\left( m_1, m_2, z \right),
\end{multline}
where
\begin{multline}
\alpha\left( \Theta \right) = \int \dd \chi_\mathrm{eff} \, \dd m_1 \, \dd m_2 \, \dd z \, P_\mathrm{det}\left( \chi_\mathrm{eff}, m_1, m_2, z \right) \\ \times p\left( \chi_\mathrm{eff} \mid \Theta, m_1, m_2 \right) p\left( m_1, m_2, z \right)
\end{multline}
is the population-averaged detection probability.

\begin{figure}
\includegraphics[width=1.0\linewidth]{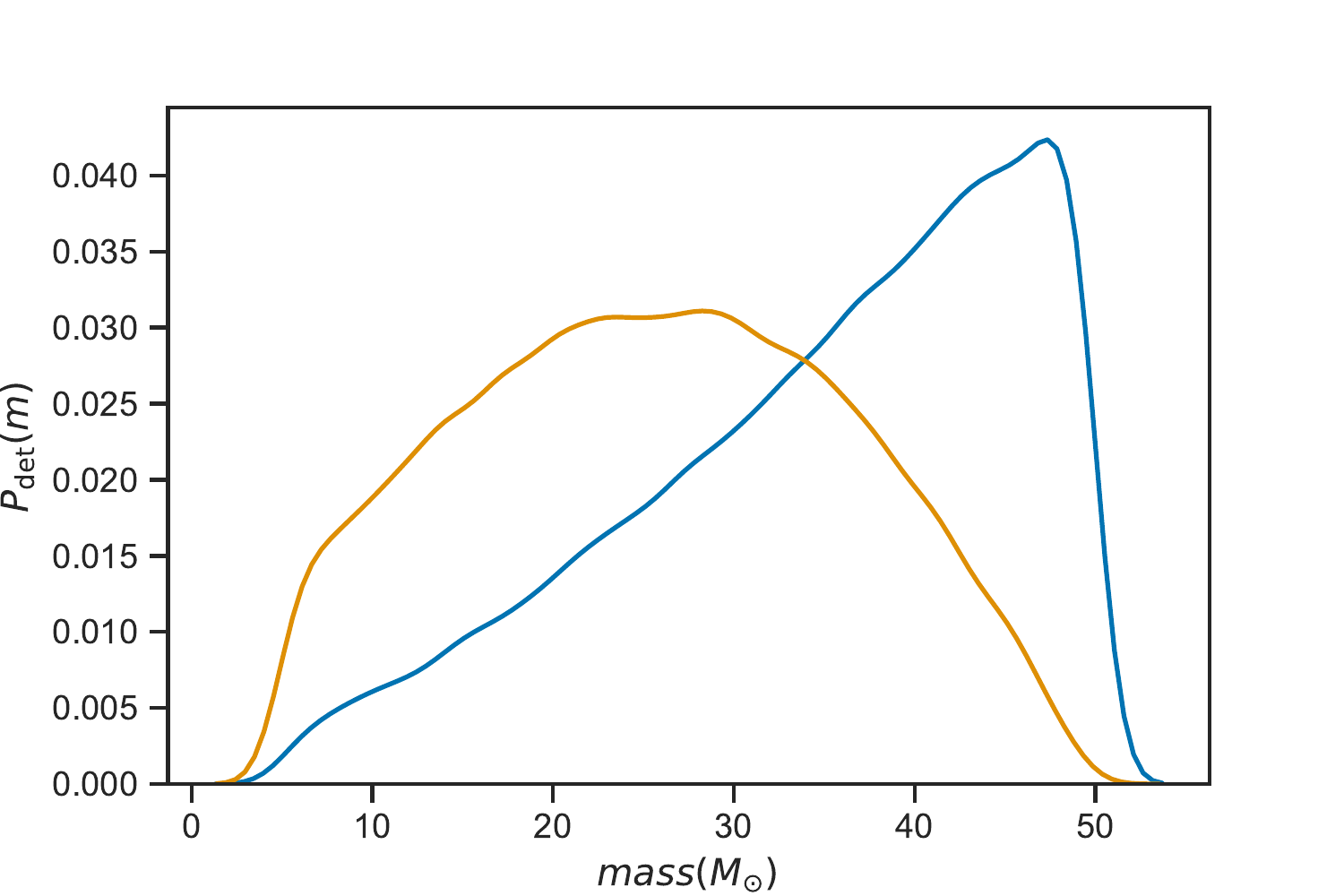}
\includegraphics[width=1.0\linewidth]{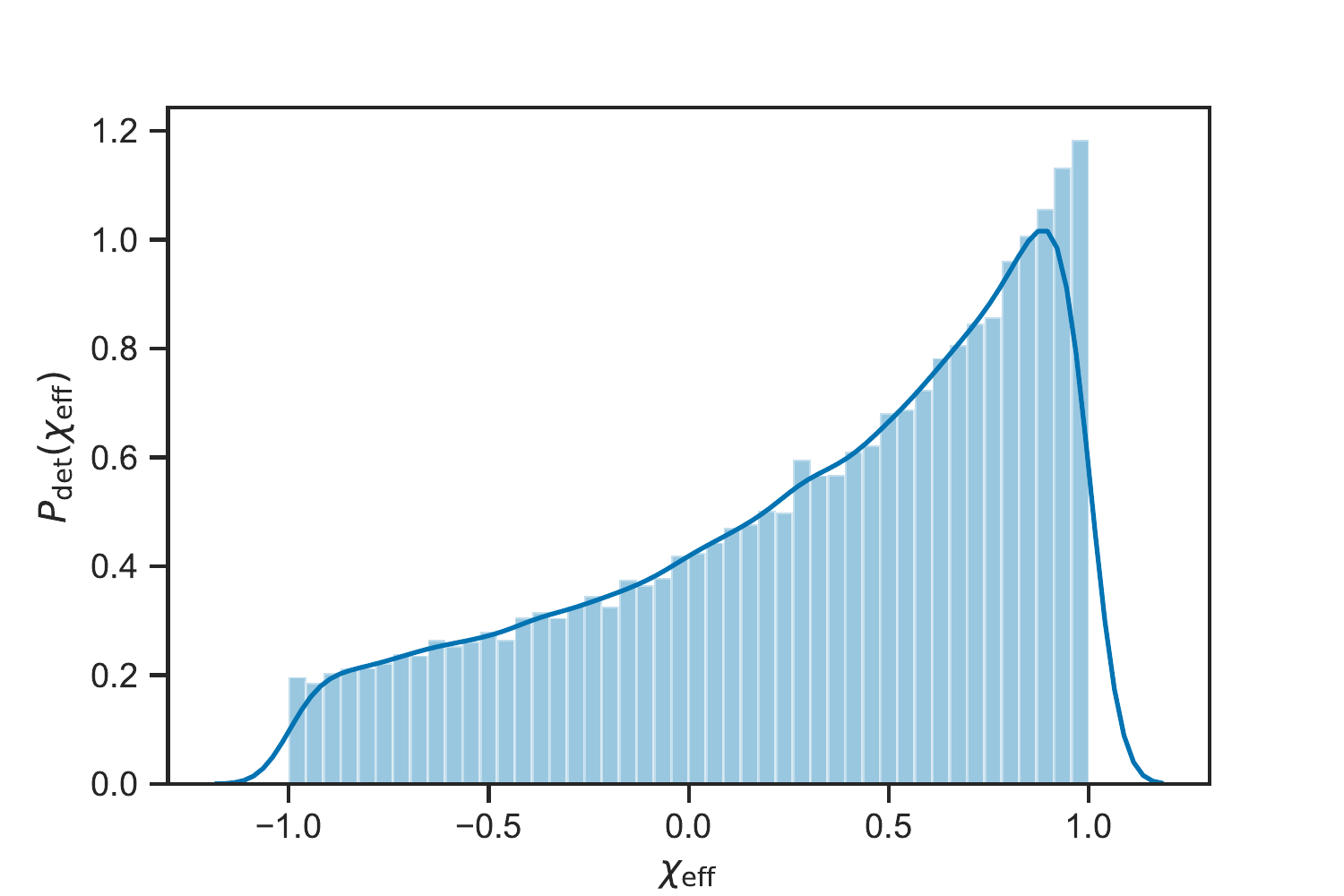}
\caption{\emph{Top panel:} LAL Detection probability for $m_1$, and $m_2$ shown in Blue and orange respectively.
\emph{Bottom panel:} shows the detection probability for \chieff for a population of BBHs with $P(m_1)\propto m_1^{-1}$, and uniform distribution in $m_2$ between 5 and 50 solar mass.
The \chieff distribution follows LAL prior.}
\label{f:p_def}
\end{figure}

Here we model the detection process semi-analytically, using a method similar to
the one described in \citet{2016ApJS..227...14A}, but with a three-detector
network (two Advanced LIGO, one Advanced Virgo, all assumed to operate with
``early high-sensitivity'' noise \citep{2018LRR....21....3A}) and a
correspondingly enhanced (noisy) SNR threshold of $\rho > 8 \sqrt{2} \simeq
11.3$. For all these calculations we use the IMRPHENOMPV2 waveform family \citep{2014PhRvL.113o1101H}. The detection probabilities produced for \chieff, $m_1$, and $m_2$ from
our analytic model are shown in Figure \ref{f:p_def}.

\begin{figure}
\includegraphics[width=1.0\linewidth]{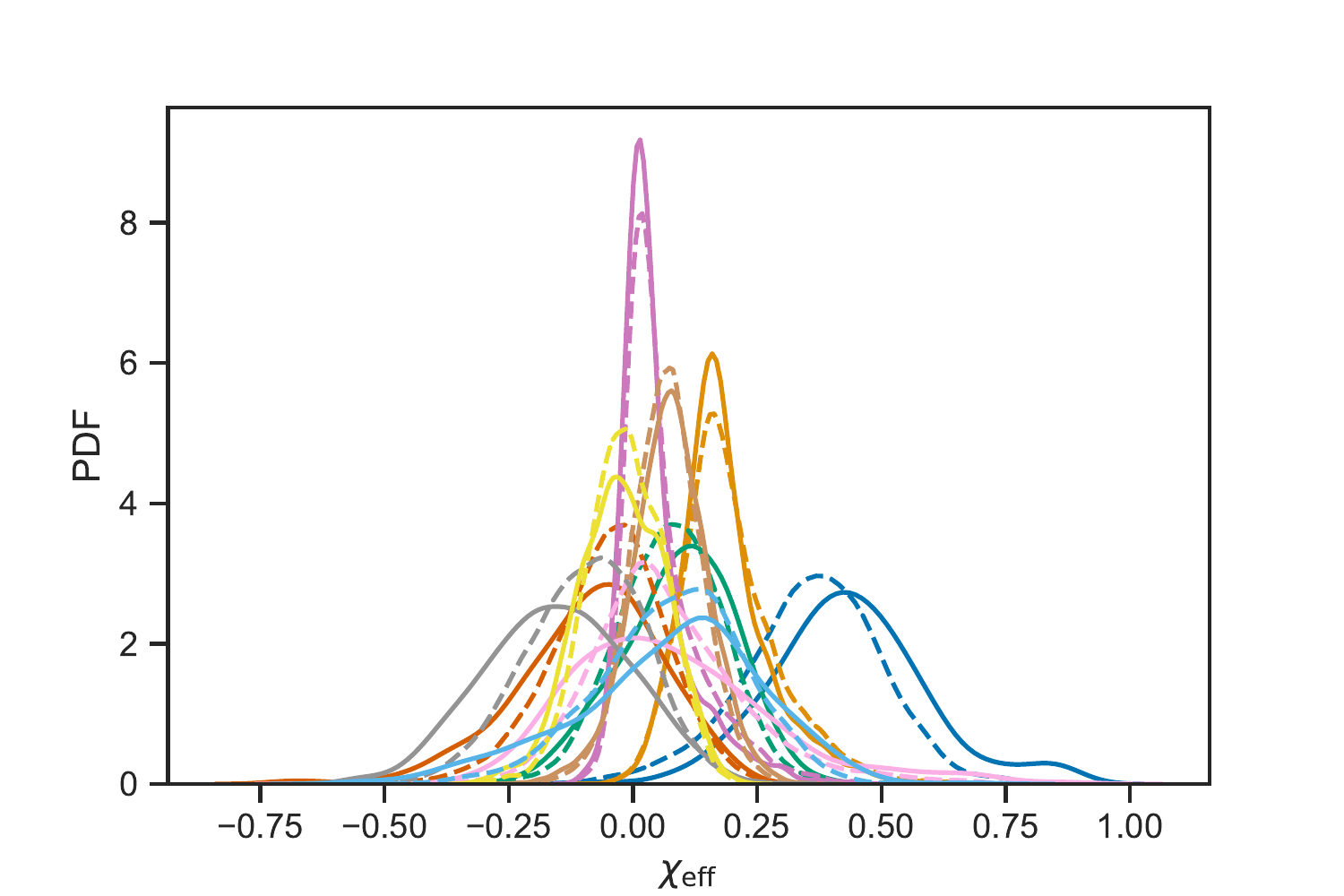}
\caption{ The posterior on \chieff from GWTC-1 (dashed lines), and the
likelihood (solid lines) after marginalizing over our assumed population in mass
and redshift (Eqs. \eqref{eq:m1-pop}, \eqref{eq:m2-pop}, and \eqref{eq:z-pop})
for the ten BBHs from GWTC-1. Removing the GWTC-1 prior on effective spin makes
the distributions move away from zero \citep{2017PhRvL.119y1103V}, but the
effect is not severe. \label{f:chi_eff_likelihood}}
\end{figure}

\begin{figure*}
\includegraphics[width=1\columnwidth]{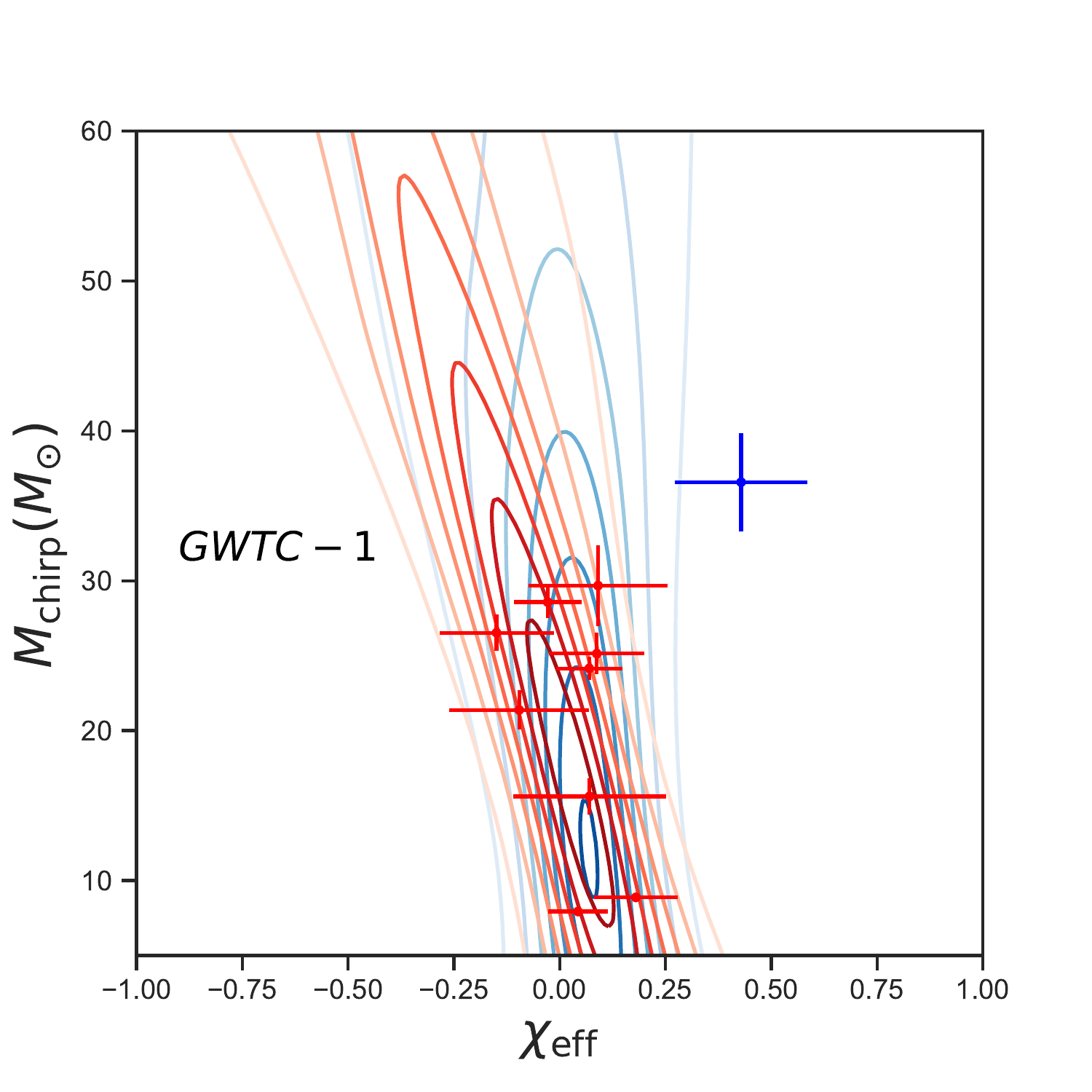}
\includegraphics[width=1\columnwidth]{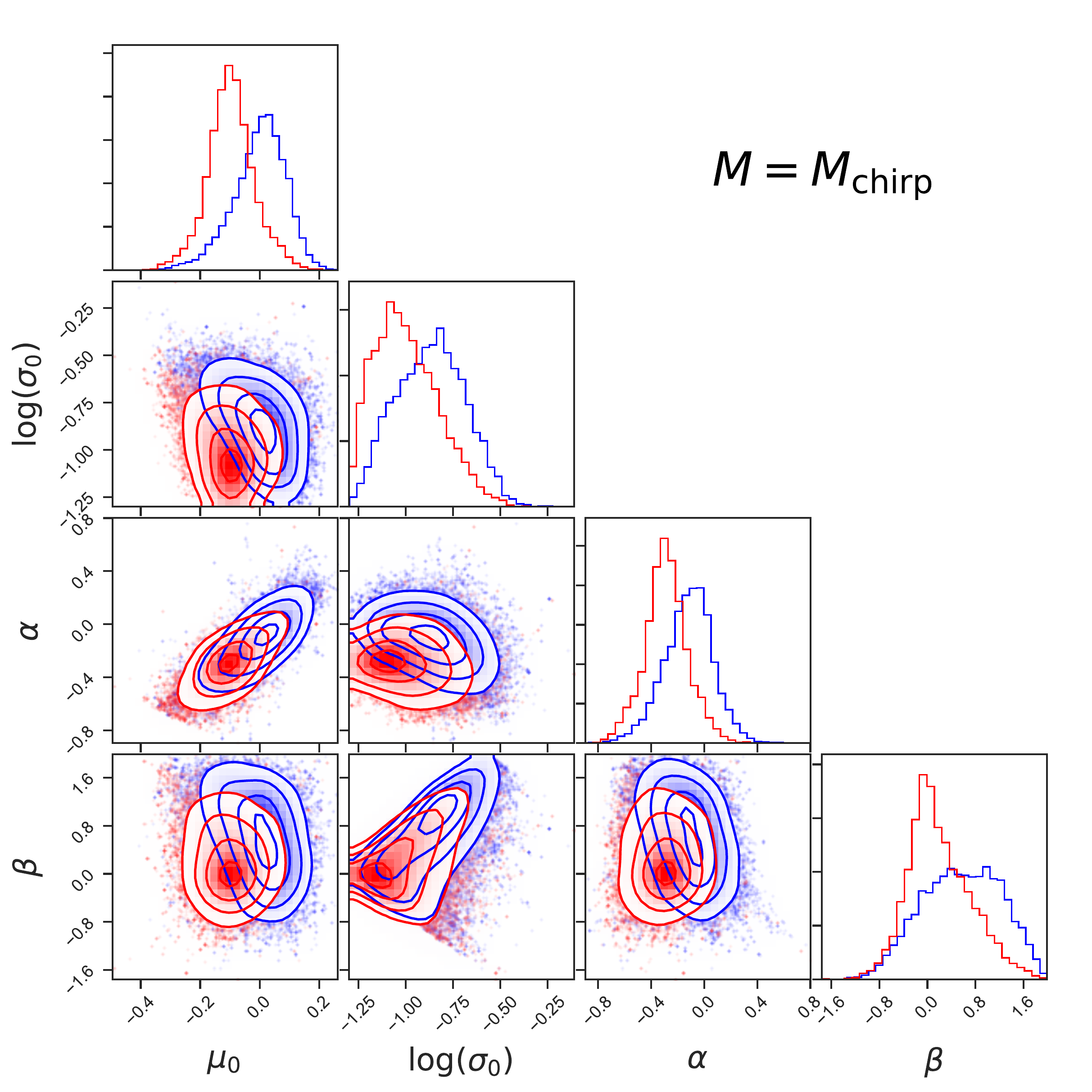}
\caption{\emph{Left panel:} shows the density $p\left( \chieff \mid
m_\mathrm{chirp}, d_{GW} \right)$ in the \chieff-$m_\mathrm{chirp}$ plane for
the 10 LIGO/Virgo BBHs in O1/O2 runs. The crosses show the $1\sigma$ error in
mass and \chieff of the ten LIGO/Virgo BBHs in GWTC-1. The blue cross represents
GW170729 which has the highest \chieff while having the highest FAR among the 10
BBHs. The red contours show the posterior PDF in the \chieff-$m_1$ plane without
considering GW170729, while the blue contours show the result when all of the
ten LIGO events are analyzed simultaneously. \emph{Right panel}: shows the
posterior PDF of the four parameters in our model used to describe the
mass-effective spin relation. A negative trend with mass ($\alpha<0$) and a
positive trend of \chieff dispersion with mass ($\beta>0$) is favored at 95\%
and 60\% respectively for the case of ignoring GW170729, and 74\%, 78\% is
favored in case of analyzing all the ten LIGO events. Our inferred values of
$\mu_0$ and $\sigma_0$ (the mean and dispersion of the $\chieff$ population at
$m_\mathrm{chirp} = 30 \, M_\odot$) are broadly consistent with the population
mean and dispersion found in \citet{Roulet:2019js}.} \label{f:posterior_M_chirp}
\end{figure*}

\begin{figure*}
\includegraphics[width=1\columnwidth]{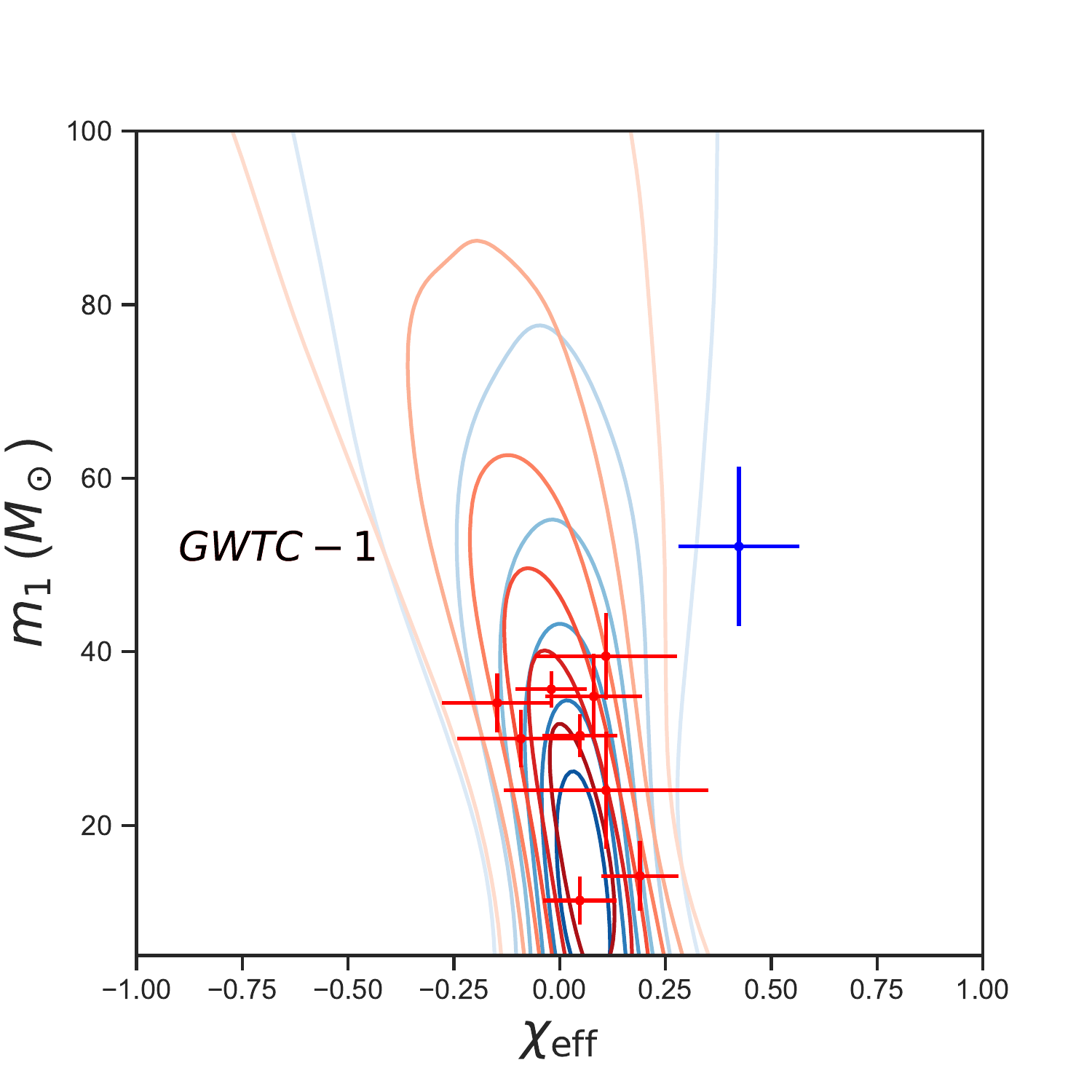}
\includegraphics[width=1\columnwidth]{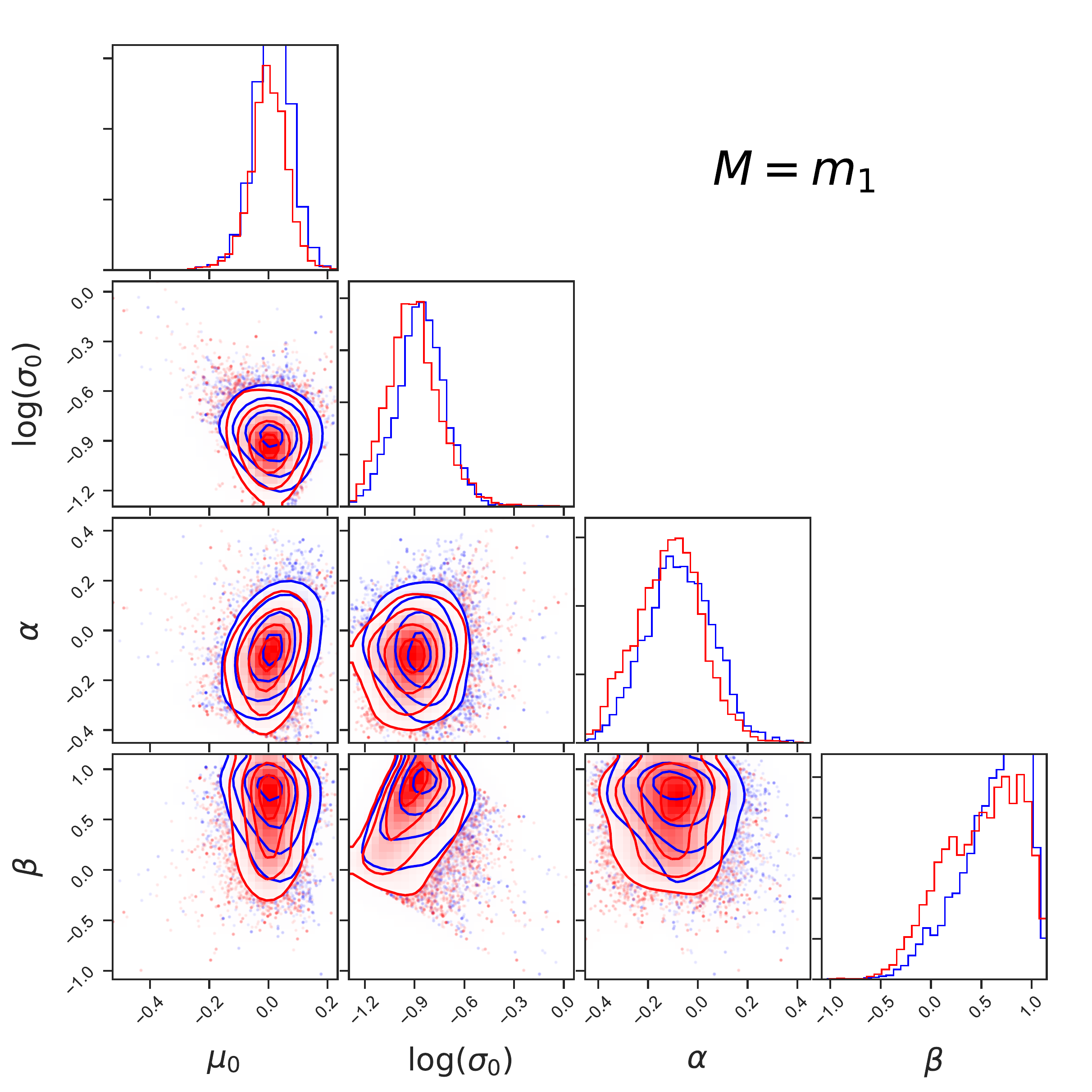}
\caption{Same as Figure \ref{f:posterior_M_chirp}, but using primary mass:
$p\left( \chieff \mid m_1\right)$, analyzing the GWTC-1 catalog. Same qualitative trends are observed: a negative
trend with mass ($\alpha<0$) and a positive trend of \chieff dispersion with
mass ($\beta>0$) is favored at 80\% and 90\% respectively for the case of
ignoring GW170729, and 70\%, 95\% is favored in case of analyzing all the ten
LIGO events. \label{f:posterior_m1}}
\end{figure*}

The integral in the product can be approximated as a weighted sum over samples drawn from the likelihood function $p\left( d_{GW}^{(i)} \mid \chi_\mathrm{eff}^{(i)}, m^{(i)} \right)$:
\begin{multline}
\int \dd \chi_\mathrm{eff} \, \dd m_1 \, \dd m_2 \, \dd z \, p\left( d_{GW}^{(i)} \mid \chi_\mathrm{eff} , m_1, m_2, z\right) \\ \times p\left(\chi_\mathrm{eff} \mid \Theta,m_1, m_2 \right) p\left( m_1, m_2, z \right) \\
 \propto \frac{1}{N_\mathrm{samp}} \sum_{j=1}^{N_\mathrm{samp}} p\left( \chi_\mathrm{eff}^{(j)} \mid \Theta,m^{(j)}_1, m^{(j)}_2 \right) p\left( m_1, m_2, z \right).
\end{multline}

Similarly, $\alpha$ can be approximated as a weighted sum over samples drawn
from a canonical distribution, $p_\mathrm{draw}\left( \chieff, m_1, m_2, z
\right)$, and ``detected'' by our synthetic pipeline
\citep{2019RNAAS...3...66F}:
\begin{equation}
\alpha\left( \Theta \right) \propto \sum_{k=1}^{N_\mathrm{detected}} \frac{p\left( \chieff \mid \Theta, m_1, m_2 \right) p\left( m_1, m_2, z \right)}{p_\mathrm{draw} \left( \chieff, m_1, m_2, z\right)}.
\end{equation}

We re-weight the posterior samples of \chieff provided by the LVC for each of
the BBH systems in GWTC-1 \citep{2019PhRvX...9c1040A} by the inverse of the
\texttt{LALInference} prior \citep{2015PhRvD..91d2003V} to draw samples from the
likelihood function. The default prior from \texttt{LALInference} assumes a
uniform mass distribution for $m_1$ and $m_2$ in the detector frame, uniform
distribution for the magnitude of the spin parameter for each black hole, and
uniform prior on the $\cos(\theta)$ from -1 to 1, and a prior on the luminosity
distance proportional to $d_L^2$.  Figure \ref{f:chi_eff_likelihood} shows the
difference between the posterior samples from the GWTC-1 catalog and the
likelihood function for $\chieff$ marginalized over our assumed mass and
redshift distribution (Eqs. \eqref{eq:m1-pop}, \eqref{eq:m2-pop}, and
\eqref{eq:z-pop}).

\begin{figure}
\includegraphics[width=1\columnwidth]{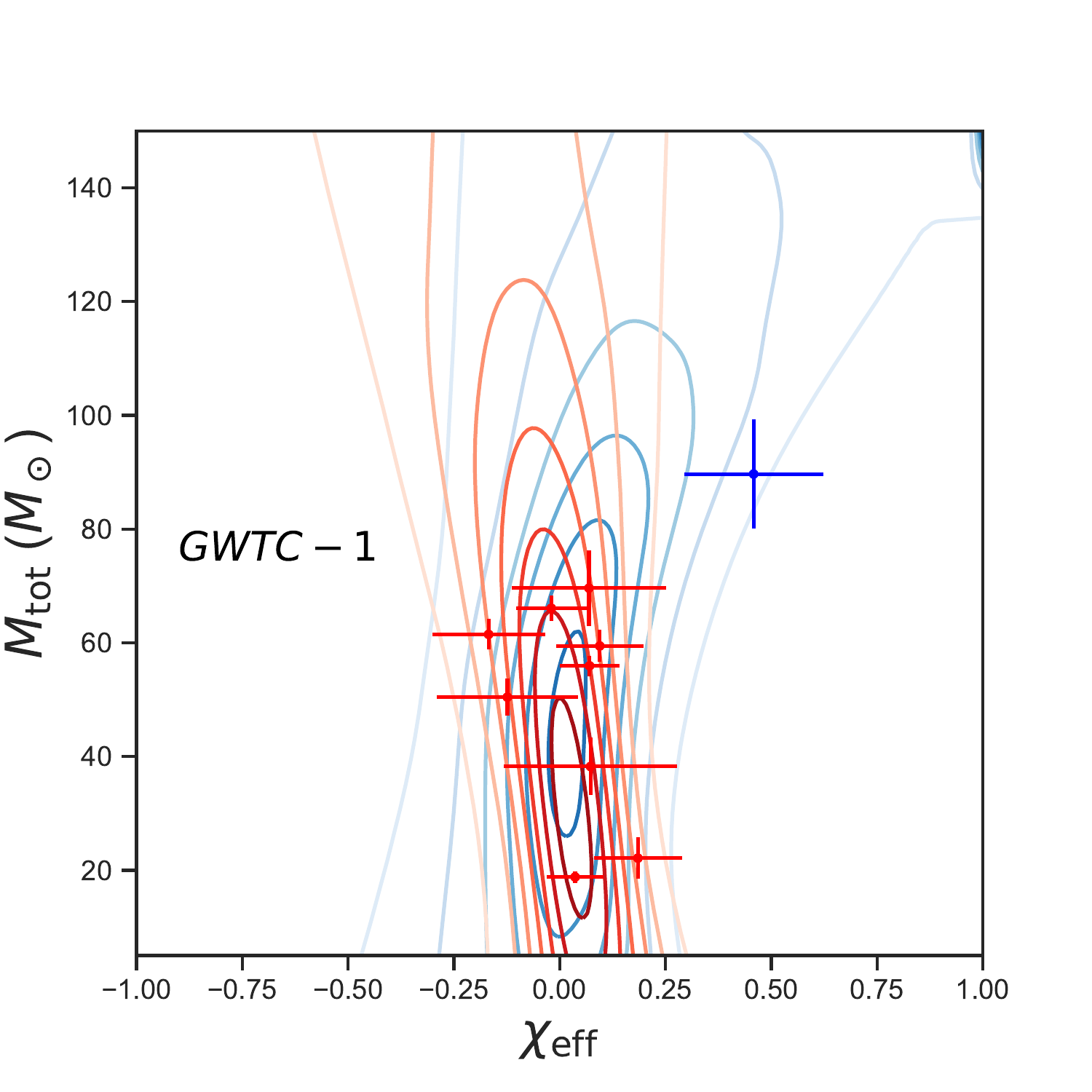}
\includegraphics[width=1\columnwidth]{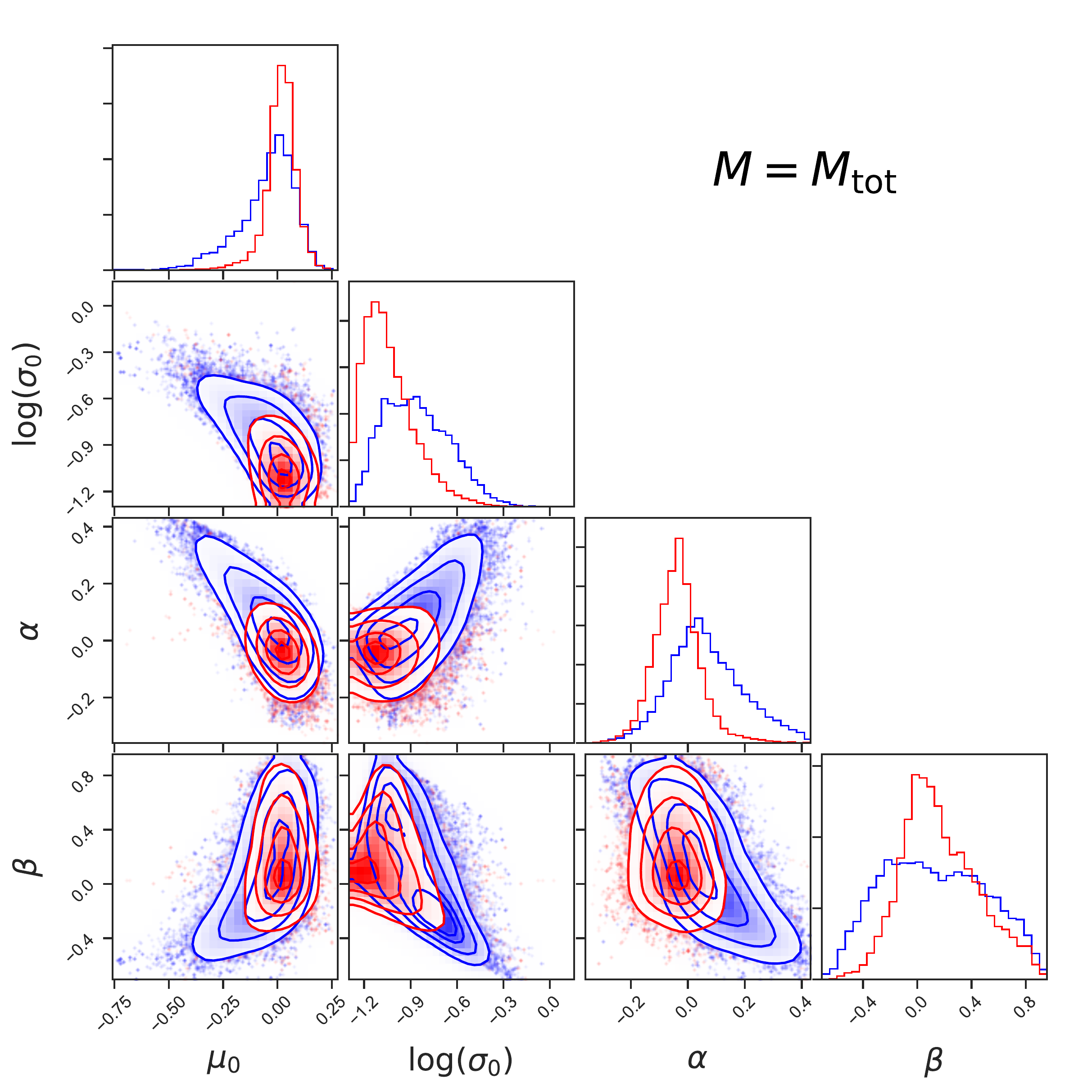}
\caption{Same as Figures \ref{f:posterior_M_chirp} and \ref{f:posterior_m1}, but
using total mass: $p\left( \chieff \mid M_\mathrm{tot}, d_{GW} \right)$
analyzing the GWTC-1 catalog. The results become rather sensitive to the inclusion of GW170729 data point. 
A negative trend with mass ($\alpha<0$) and a
positive trend of \chieff dispersion with mass ($\beta>0$) is favored at 30\%
and 60\% respectively for the case of ignoring GW170729, and 70\%, 70\% is
favored in case of analyzing all the ten LIGO events. \label{f:posterior_M_tot}}
\end{figure}

We apply a flat prior on the population parameters $\Theta$, and sample from
their posterior distribution given the catalog data (i.e. we draw samples of
$\Theta$ proportional to the function in Eq. \eqref{eq:catalog-likelihood})
using the \texttt{emcee} stochastic sampler \citep{ForemanMackey:2013io} with
various choices of mass parameter controlling the $\chieff$ distribution.  Our
results are summarized in Figures \ref{f:posterior_M_chirp},
\ref{f:posterior_m1}, and \ref{f:posterior_M_tot}.

The left panel of Figure \ref{f:posterior_M_chirp} shows a contour plot of the
posterior $\chieff$ distribution in the \chieff-$m_\mathrm{chirp}$ plane after
analyzing the GWTC-1 BBHs and marginalizing over $\Theta$:
\begin{equation}
  p\left( \chieff \mid d_{GW} \right) = \int \dd \Theta \, p\left( \chieff \mid \Theta \right) p\left( \Theta \mid d_{GW} \right).
\end{equation}
The crosses show the $1\sigma$ error in chirp mass and \chieff of the ten LIGO
BBHs. The blue cross represents GW170729 which has the highest observed \chieff
among the 10 BBH systems. The red contours show the posterior PDF, $p\left(
\chieff \mid m_\mathrm{chirp}, d_{GW}\right)$ without considering GW170729,
while the blue contours show the result when all of the ten LIGO events are
analyzed. We have singled out GW170729 since it has the highest significant false
alarm rate (FAR) of 0.18 / year in the GstLAL pipeline \citep{Abbottetal:2018vb}
which nearly matches the threshold of 0.1 / year, while the other BBH systems
have significantly lower FARs ($<10^{-3}$).

The right panel of Figure \ref{f:posterior_M_chirp} shows the posterior PDF of
the four parameters in our model used to describe the \chieff-mass relation. A
negative trend with mass ($\alpha<0$) and a positive trend of \chieff dispersion
with mass ($\beta>0$) are favored at 95\% and 60\% respectively for the case of
ignoring GW170729, and 74\%, 78\% when analyzing all ten LIGO events.

Figure \ref{f:posterior_m1} shows the same result but when the mass variable is
taken to be the primary mass of the system. When the mass scale is the primary
mass of the BBHs, a negative trend with mass ($\alpha<0$) and a positive trend
of \chieff dispersion with mass ($\beta>0$) is favored at 80\% and 90\%
respectively for the case of ignoring GW170729, and 70\%, 95\% in case of
analyzing all the ten LIGO events.

Figure \ref{f:posterior_M_tot} shows the same result but when the mass is taken
to be the total mass of the system. When the mass scale is the total mass of the
BBHs, a negative trend with mass ($\alpha<0$) and a positive trend of \chieff
dispersion with mass ($\beta>0$) is favored at 30\% and 60\% respectively for
the case of ignoring GW170729, and 70\%, 70\% is favored in case of analyzing
all the ten LIGO events.

Although the choice of mass scale somewhat affects the confidence by which a
trend with mass for either the mean \chieff or its dispersion could be detected,
the data suggests that if there is a trend, it is a negative trend with mass for
the mean \chieff and a positive trend for its dispersion. Dynamical assembly
alone can account for the larger dispersion in \chieff with mass, however the
negative trend of the mean \chieff with mass can not be accounted for based on
dynamical assembly. Field formation of the BBHs on the other hand can
potentially explain the negative trend with mass (through a combination of
tidally spun up systems at low masses and massive BBHs with about zero effective
spin at birth), however, the increase of dispersion with mass would be hard to
accommodate based on field evolution alone.  Thus we suggest that the observed
trends in $\chieff$ with mass could be indicating the operation of \emph{both}
formation channels in the GWTC-1 observations.

\section{Summary \& Conclusion}\label{sec:summary}
A given formation channel for a BBH system would predict a certain distribution in \chieff-mass plane for the final merger event that LIGO/Virgo would observe.
Field binaries tend to predict a banana shaped region that encompass massive systems with negligible \chieff magnitude or low mass systems with positive \chieff magnitudes if the angular momentum transport is weak.
The distribution could be combination of three main mechanism : (i) formation with negligible spin at all masses due to efficient AM transport, (ii) tidal spin of the binaries in which the second born
compact object forms from a tidally spun up progenitor in a close orbit with another BH, (iii) formation with moderate efficiency of AM transport while including feedback from disk formation after the core
of the progenitor star has collapsed and formed a BH.

On the other hand, dynamical assembly of BBHs in dense stellar clusters leads to a symmetric distribution of BBHs in \chieff at all masses with larger dispersion at higher masses.
The increase of dispersion is due to a random walk in \chieff-mass that higher generation BHs follow.

In this study we find a tentative negative correlation between \chieff and chirp
mass for the ten LIGO/Virgo BBHs with $\sim 75\%$ confidence. The negative
correlation could in principle be explained by field formation alone. However,
standard field formation consistent with the observed small effective spins at
low mass would predict that the dispersion should decrease with mass, the
opposite of what the data suggests. We find that the dispersion in \chieff grows
with mass with  80\% confidence. These trends are consistent with a combined
channel of dynamically assembled BBHs that provide the positive trend of
dispersion with mass, and a field formation channel that provides the negative
mean trend with mass could explain our findings.

Given the public alerts released by Advanced LIGO and Virgo\footnote{See
\url{https://emfollow.docs.ligo.org/userguide/}.} in the first half of the third
observing run, the O3a catalog should contain $\sim 30$ BBH mergers; therefore,
the statistical uncertainties on our parameters $\Theta$ should decrease by
about a factor of $\sqrt{N_{O3a}/N_{O2}} \sim 2$ on incorporating the O3a
catalog; additional detections in O3b
\citep{2019PhRvX...9c1040A,2018LRR....21....3A} should produce a further factor
of $\sim \sqrt{2}$ reduction in uncertainty. Thus we can expect the O3 catalog
of BBH systems to confidently confirm or overturn the trends observed here.

\software{
Numpy \citep{numpy},
Scipy \citep{scipy},
IPython \citep{IPython},
Matplotlib \citep{Matplotlib},
astropy \citep{astropy:2013,astropy:2018},
PyStan \citep{Stan,PyStan},
Seaborn \citep{Seaborn},
Arviz \citep{Arviz},
emcee \citep{ForemanMackey:2013io},
corner \citep{corner}
}

\acknowledgements MTS is grateful to the Center for Computational Astrophysics
for hospitality during the course of this work.  ER-R and MTS thank the
Heising-Simons Foundation, the Danish National Research Foundation (DNRF132) and
NSF (AST-1911206 and AST-1852393) for support.  The authors thank the LIGO
Scientific Collaboration for access to public data and gratefully acknowledge
the support of the United States National Science Foundation (NSF) for the
construction and operation of the LIGO Laboratory and Advanced LIGO as well as
the Science and Technology Facilities Council (STFC) of the United Kingdom, and
the Max Planck-Society (MPS) for support of the construction of Advanced LIGO.
Additional support for Advanced LIGO was provided by the Australian Research
Council. We are thankful to Daniel Wysocki, Richard O'shaughnessy, and Jim Fuller for useful comments on the earlier version of this work.

\bibliographystyle{apj}
\bibliography{the_entire_lib.bib}

\end{document}